\documentclass[preprint,aps]{revtex4}
\usepackage{graphicx}
\input epsf
\begin{document}

\title{\bf  A dark energy model resulting from a Ricci symmetry revisited}
\author{R. G. Vishwakarma\footnote{Electronic Address: rvishwa@mate.reduaz.mx, rvishwak@ictp.it}}
\address{Unidad Acad$\acute{e}$mica de Matem$\acute{a}$ticas\\
 Universidad Aut$\acute{o}$noma de Zacatecas\\
 C.P. 98068, Zacatecas, ZAC.\\
 Mexico}


\begin{abstract}
Observations of supernovae of type Ia require dark energy (some unknown exotic \emph{`matter'} of
negative pressure) to explain their unexpected faintness. Whereas the simplest and most favoured
candidate of dark energy, the Einsteinian cosmological constant, is about 120 orders of magnitude
smaller than the theoretically predicted value. Motivated by this problem, a number of models of dynamically decaying dark energy have been proposed by considering different phenomenological laws 
or potentials of the scalar field, which are more or less ad-hoc.
However, it is more advisable to consider the symmetry properties of spacetime rather than the ad-hoc assumptions. 

In this view,  we consider a model of Robertson-Walker cosmology emerging from a Ricci  
symmetry which provides consistently an evolving dark energy. We test the model for the recent supernovae Ia data, as well as, the
ultracompact
radio sources data compiled by Jackson and Dodgson. The model fits the data very well.

\bigskip
\noindent
{\it Subject heading:} cosmology: theory - cosmology: observations -
cosmological constant.

\noindent
{\bf Key words:} cosmology: theory - cosmology: observations -
cosmological constant.

\noindent
PACS: 98.80.-k,~98.80.Es,~98.90.+s
\end{abstract}

\maketitle


\section{Introduction}

\noindent
It is generally believed that the present expansion of the universe is accelerating
due to the presence of some unknown cosmic `matter' with negative pressure generally
termed as {\it dark energy}. The simplest and the most favoured candidate of dark 
energy is Einstein's cosmological constant $\Lambda$, which is though plagued with 
horrible fine tuning problems. This happens due to the presence of two values of $\Lambda$
differing from each other by some 120 orders of magnitudes: the magnitude of $\Lambda$ at 
the beginning of inflation and the value given by the present-day observations. 
This has led a number of cosmologists to consider models 
of evolving dark energy, by either proposing different phenomenological laws or by considering
different potentials of the scalar fields which are more
or less ad-hoc. The precise mechanism of the evolution of dark energy, which could be
required by some symmetry principle, is not yet known. Indeed, since  
the basic motivation in these models is to understand the present-day
smallness of $\Lambda$, they do not provide any natural relation between
the magnitude of $\Lambda$ at the beginning of inflation and the
present-day observational value.

It would be worth while to investigate some symmetry principles
behind the problem crying out for the evolution of dark energy and thus
develop a more realistic and fundamental model for dark energy. 
Moreover it is always reasonable to consider symmetry
properties of spacetime rather than considering ad-hoc assumptions. 
In this view, we have discovered \cite{curr, sattar, vishwa_grg} a model resulting from a
contracted Ricci-collineation which,
apart from having interesting conservation properties, does
provide a dynamical law for decaying $\Lambda$. The physical properties of the model 
have been discussed in \cite{curr, sattar, vishwa_grg} and it was found that the model had a credible 
magnitude-redshift ($m$-$z$) relation for the
observations of supernovae (SNe) of type Ia from Perlmutter 
et al. \cite{perl}. Since then many observations of SNe Ia have been made, quite many at higher redshifts, by the Hubble Space Telescope. It would be worthwhile to examine how well (or badly) the model fits the new data.

Like the luminosity of a standard candle, the angular size $\Theta$ of a standard measuring rod changes with 
its redshift $z$ in a manner that depends upon the parameters of the model. 
Hence the $\Theta$-$z$ 
relation is also proposed as a potential test for 
cosmological models by Hoyle \cite{hoyle}. Therefore, it would also be 
worthwhile
to examine the $\Theta$-$z$ relation in this model for the
dataset of Jackson and Dodgson \cite{JacDod}, which is a trustworthy compilation
of ultracompact radio sources and has been already used to test different cosmological
models \cite{JacDod, BanNar}.
In the following section we describe the model in brief for ready reference and to derive
 the observational relations. More details can be found in \cite{vishwa_grg}.

\section{The model}

\noindent
As we are going to consider a dynamical $\Lambda$ in Einstein's theory, it 
would be worthwhile to mention a general result which holds irrespective of the dynamics 
of $\Lambda$: the empty spacetime of de Sitter cannot be a solution of general relativity 
with a dynamical $\Lambda(t)$ \cite{vishwacqg19}. This follows from the divergence of the field 
 equation
$[R^{ij}-\frac{1}{2} R g^{ij}]_{;j}=0=\left[T^{ij}_{(\rm m)}-g^{ij}\Lambda(t)/8\pi G\right]_{;j}$. Obviously a solution with a dynamical $\Lambda(t)$ is 
possible only if $T^{ij}_{(\rm m)}\neq0$ (and $T^{ij}_{(\rm m);j}\neq0$).
We assume that the universe is homogeneous and isotropic, represented by the
Robertson-Walker (RW) metric, and its dynamics is given by the Einstein field 
equations 
\begin{equation}
-\frac{\ddot S}{S}=\frac{4\pi
G}{3}\left(\rho _{\mbox{{\scriptsize tot}}}+3p _{\mbox{{\scriptsize tot}}}\right),\label{eq:Rchaudhuri}
\end{equation}

\begin{equation}
\frac{\dot S^2}{S^2}+\frac{k}{S^2}=\frac{8\pi
G}{3}\rho _{\mbox{{\scriptsize tot}}},\label{eq:Freedman}
\end{equation}
where $S$ and $k$ are respectively the scale factor and curvature index
appearing in the RW metric;
and $\rho _{\mbox{{\scriptsize tot}}}=\rho_{\rm m}+
\rho_{\Lambda}\equiv \rho_{\rm m}+\Lambda(t)/8\pi G$,
$p _{\mbox{{\scriptsize tot}}}=p_{\rm m}+p_{\Lambda}=p_{\rm m}-
\rho_{\Lambda}$, with $\Lambda$ representing the cosmological term.

It is well known that collineations of Ricci tensor ($R_{ij}$) have interesting symmetry
properties and lead to useful conservation laws in general relativity \cite{ricci, curr, sattar, 
vishwa_grg}. 
The Ricci collineation along a vector $\eta_i$ is defined by vanishing Lie-derivative of $R_{ij}$
 along $\eta_i$: ${\cal L}_\eta R_{ij}=0$.
It has been shown \cite{curr, sattar} that in general relativity the contracted 
Ricci-collineation along the fluid flow $u^i$ (normalized velocity 4-vector), i.e.,
$g^{ij}{\cal L}_u R_{ij}=0$,
leads to the conservation of generalized {\it momentum density}:
\begin{equation}
\{(\rho _{\mbox{{\scriptsize tot}}} + 3p _{\mbox{{\scriptsize tot}}})u^j\}_{;j}=0.\label{eq:con1}
\end{equation}
For the RW metric,
the conservation law (\ref{eq:con1}) leads to the conservation of the
total active gravitational mass of a comoving sphere of radius $S$:
\begin{equation}
(\rho _{\mbox{{\scriptsize tot}}} + 3p _{\mbox{{\scriptsize tot}}})S^3 =
\mbox{constant} = A ~ ~ \mbox{(say)}\label{eq:con2}
\end{equation}
(A detailed discussion elaborating on the meaning of (\ref{eq:con2}) has been done in 
\cite{vishwa_grg}. Consequences of the resulting models for the case $\Lambda=0$ have been discussed in \cite{curr}.) 
As there are 4 unknowns $\rho_{\rm m}$, $p_{\rm m}$, $\rho_\Lambda$ ($=\Lambda(t)/8\pi G$) and the scale factor $S$, the equations (\ref{eq:Rchaudhuri}), (\ref{eq:Freedman}) and (\ref{eq:con2}), together with  the usual barotropic equation of state $p_{\rm m}/\rho_{\rm m}=$ constant $=w$ ($0\leq w\leq 1$) for the matter source, provide a unique solution of the model.
By using (\ref{eq:con2}) in the Raychaudhuri equation
(\ref{eq:Rchaudhuri}) and integrating the resulting equation, we get
\begin{equation}
\frac{\dot S^2}{S^2}=\frac{8\pi GA}{3S^3}+\frac{B}{S^2},\label{eq:H}                                           
\end{equation}
supplying the dynamics of the scale factor, where $B$ is a constant of integration. Equations (\ref{eq:Freedman}), (\ref{eq:con2}) and (\ref{eq:H}) supply the unique dynamics of dark energy density $\rho_\Lambda$ and $\rho_{\rm m}$  as
\begin{equation}
\rho_\Lambda=\frac{1}{(1+w)}\left[\frac{w 
A}{S^3}+\frac{(1+3w)(B+k)}{8\pi G S^2}\right],\label{eq:lam}
\end{equation}
\begin{equation}
\rho_{\rm m}=\frac{1}{(1+w)}\left[\frac{A}{S^3}+\frac{(B+k)}{4\pi G S^2}\right].\label{eq:rhom}                      
\end{equation}
It must be noted that we have not assumed the conservation of the matter source
 which is usually done through the additional assumption of no interaction 
(minimal coupling) between different 
source fields (except for the case with a constant $\Lambda$ which is consistent with the idea of minimal coupling), which though seems ad-hoc and nothing more than a simplifying 
assumption. On the contrary, interaction is more natural and is a fundamental 
principle. Let us recall that the only constraint on the
source terms, which is imposed by Einstein's equation (through the Bianchi identities), is the conservation of the sum of all the 
energy-momentum tensors, individually they are not conserved:
$[R^{ij}-\frac{1}{2} R g^{ij}]_{;j}=0=\left[T^{ij}_{(\rm m)}+T^{ij}_{(\Lambda)}+T^{ij}_{(\phi)}+ ...\right]_{;j}$, implying creation or annihilation for the
case $\Lambda(t)$.

It may be noted that the evolution of $\rho_\Lambda$, as given by equation (\ref{eq:lam}), is 
a function of the equation of state of matter. This may be regarded as a kind of generalization of 
the ansatz $\Lambda \propto S^{-2}$ proposed by many authors \cite{s-2}, which is obtained in the
present model in the present phase of evolution of the universe ($w=0$).

\medskip
In order to study the $\Theta$-$z$ and the $m$-$z$ relations in the model, 
we can rewrite 
equations (\ref{eq:H}-\ref{eq:rhom}) by specifying the constants $A$ and $B$ in terms
of the cosmological parameters in the present phase of evolution:
$A=(\Omega_0-2\Omega_{\Lambda 0}) 3S_0^3 H_0^2/8\pi G$,~~
$B=(2\Omega_{\Lambda0}-\Omega_0+1)H_0^2 S_0^2$, ~~giving
\begin{equation}
\Omega_\Lambda (z)=\Omega_{\Lambda0}(1+z)^2\frac{H_0^2}{H^2(z)},
\end{equation}

\begin{equation}
\Omega_{\rm m} (z)=[(\Omega_{\rm m0}-2 \Omega_{\Lambda0})(1+z)^3+2\Omega_{\Lambda0}(1+z)^2]\frac{H_0^2}{H^2(z)},
\end{equation}
where
\begin{equation}
H (z)=H_0[(\Omega_{\rm m0}-2 \Omega_{\Lambda0})(1+z)^3-(\Omega_{\rm m0}-2 \Omega_{\Lambda0}-1)(1+z)^2]^{1/2}.\label{eq:hubble}
\end{equation}
Here $\Omega_i$ are, as usual, the energy density of different source components of the cosmological fluid in
units of the critical density $3H^2/8\pi G$  (i denoting matter 
($\rm m$), cosmological term ($\Lambda$), etc.). 
The subscript `0' denotes the value of the quantity at the present epoch. 

The angular size-redshift ($\Theta$-$z$) relation in the model is
given by Hoyle's formula 
\begin{equation}
\Theta=\frac{d(1+z)}{r S_0} ~  \mbox{radian},\label{eq:hoyle}
\end{equation}
which relates the 
apparent angular diameter $\Theta$ of the source (of redshift $z$ located at a coordinate distance 
$r$)
with its absolute angular size $d$ (presumably same for all sources).
The coordinate distance $r$ can be calculated from the RW metric according to its curvature
parameter $k$:
\begin{equation}
r=
\left\{ \begin{array}{ccl}
\vspace{0.4cm}
\sin\left(\frac{1}{S_0} \, \int_0^z \, \frac{{\rm d} z'}{H(z')} \right),& \mbox{when}& k = 1 \\
\vspace{0.4cm}
\frac{1}{S_0} \, \int_0^z \, \frac{{\rm d} z'}{H(z')}, &\mbox{when}& k = 0 \\
\sinh\left(\frac{1}{S_0} \, \int_0^z \, \frac{{\rm d} z'}{H(z')} \right),& \mbox{when}& k = -1.
\end{array}\right. \label{eq:rdist}
\end{equation}
The present value of the scale factor $S_0$, appearing in
equations (\ref{eq:hoyle}, \ref{eq:rdist}), which measures the present curvature of
spacetime, can be calculated from
\begin{equation}
S_0 = H_0^{-1} \sqrt{\frac{k}{
(\Omega_{\rm m0}+\Omega_{\Lambda0}-1)}}.\label{eq:szero}
\end{equation}
As the measured angular sizes of the radio sources in the dataset from Jackson and Dodgson
are given in units of milli arc second (mas), we rewrite equation (\ref{eq:hoyle}) as
\begin{equation}
\Theta(z; \Omega_i, d h_0)=0.0688 \times d h_0 \sqrt{\frac{
(\Omega_{\rm m0}+\Omega_{\Lambda0}-1)}{k}}~ \frac{(1+z)}{r} ~ ~ \mbox{mas},\label{eq:hoyle1}
\end{equation}
where $d$ is measured in pc (par sec)  and $h_0$ is the present value of the Hubble constant
$H_0$ measured in units of  100 Km s$^{-1}$Mpc$^{-1}$. We are now able to 
calculate the angular size $\Theta(z,\Omega_i,dh_0)$ of a radio source 
at a given redshift $z$ predicted by the model (for a given set of
the parameters $\Omega_i, dh_0$) by
using equations (\ref{eq:hubble}, \ref{eq:rdist}-\ref{eq:hoyle1}).

We also recall that the usual magnitude-redshift ($m$-$z$) relation in a 
homogeneous and isotropic model (based on the RW metric) is given by
\begin{equation}
m(z;\Omega_i,{\cal M})={\cal M} +
5 \log\{H_0 d _
{\mbox{{\scriptsize L}}}(z; \Omega_i)
\},\label{eq:mag}
\end{equation}
where $m$ is the apparent magnitude of a SN of redshift $z$ located at the 
coordinate distance $r$,  
${\cal M} \equiv M - 5 \,\log H_0 +25$,
$M$ is the absolute magnitude (presumably same for all SNe Ia),
and $d_{\rm L}$ is the luminosity distance of the SN given by
\begin{equation}
d _{\mbox{{\scriptsize L}}}(z; \Omega_i)=(1+z) S_0 r ~ ~ ~ {\rm Mpc}. \label{eq:dl}
\end{equation}
Now we can calculate the magnitude $m(z,\Omega_i,{\cal M})$ of a SN 
at a given redshift $z$ predicted by the model (for a given set of
the parameters $\Omega_i, {\cal M}$) by
using equations (\ref{eq:hubble}, \ref{eq:rdist}, \ref{eq:szero}, \ref{eq:mag}, \ref{eq:dl}). 

In order to fit the model to the observations, we calculate $\chi^2$ according
 to
\begin{equation} 
\chi^2=\sum_{i=1}^{N} \left[\frac{X^{\rm obs}_i -
X^{\rm pred}(z_i)}{\sigma_{X^{\rm obs}_{i}}}\right]^2,\label{eq:chi}
\end{equation}
where $X^{\rm obs}_i$ is the observed value of the observable, $X^{\rm pred}(z_i)$ is its predicted value at the redshift $z_i$, $\sigma_{X^{\rm obs}_{i}}$
is the uncertainty in the observed value $X^{\rm obs}_i$ and $N$ is the number of data points (or bins).
($X$ stands for $\Theta$ 
and $m$ respectively for the data of radio sources and SNe Ia.)

\begin{figure}[tbh!]
\centerline{{\epsfxsize=14cm {\epsfbox[50 250 550 550]{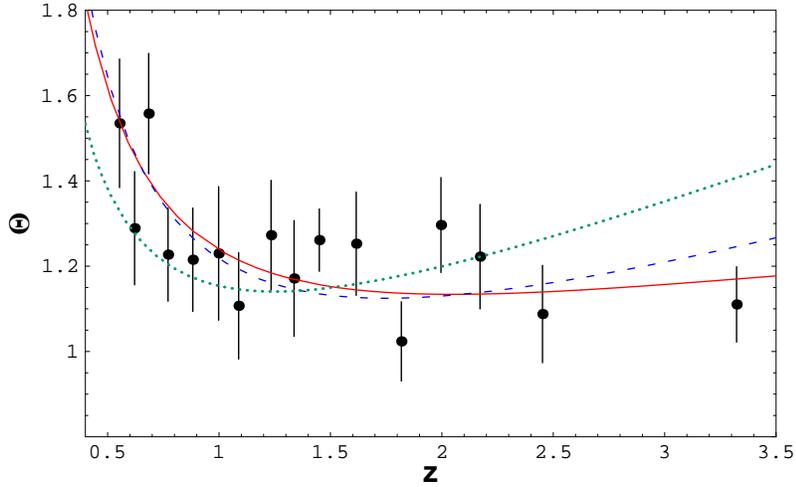}}}}
{\caption{\small Some best-fitting models to the ultracompact radio sources 
data from Jackson \& Dodgson are shown. The solid curve corresponds to the 
best-fitting dark energy model 
$\rho_{\Lambda}\propto S^{-2}$ with $\Omega_{\rm m0}=1-\Omega_{\Lambda0}=0.62$, the dashed  
curve corresponds to the best-fitting standard $\Lambda$CDM model with $\Omega_{\rm m0}=1-\Omega_{\Lambda0}=0.21$
and the dotted curve corresponds to the best-fitting Einstein-de Sitter model ($\Omega_{\rm m0}=1-\Omega_{\Lambda0}=1$). 
}}
\end{figure}

\section{Fitting the model to the radio sources data}

\noindent
We use the sample of 256 ultracompact radio sources compiled by
 Jackson and Dodgson \cite{JacDod}. 
This sample of 256 radio sources with $z$ in the range 0.5 to 3.8 was selected
by them from a bigger sample of
 337 ultracompact radio sources originally compiled by Gurvits \cite{gurvits}. 
 These sources, of angular sizes of  
the order of a few milliarcseconds (ultracompact), were measured by the {\it very long-baseline
interferometry}. The points of the sample of Jackson and Dodgson
are short-lived quasars deeply
embedded inside the galactic nuclei, which are expected to be free from
evolution on a cosmological time scale and thus comprise a set of standard
rods, at least in a statistical sense.
Jackson and Dodgson binned their sample into 16 redshift
bins, each bin containing 16 sources.
We fit the present model to this sample by calculating $\chi^2$ according to (\ref{eq:chi}) and minimize it with respect to the free
parameters $\Omega_{\rm m0}$, $\Omega_{\Lambda0}$ and $dh_0$.
The global
minimum is obtained for the values (with the constraint 
$\Omega_{\rm m0}\geq0$)

\begin{center}
\begin{tabular}{llll}
$\Omega_{\rm m0}=0.68$,& 
$\Omega_{\Lambda0}=0.45$,&
 $dh_0=6.50$& with $\chi^2=13.00$
\end{tabular}
\end{center}
\noindent
at 13 degrees of freedom (DoF). This represents a very good fit, with the 
goodness-of-fit probability $Q=44.8$\% (see the Appendix for an explanation 
of $Q$). We further note that the solution for the minimum $\chi^2$ is very 
degenerate in the parameter space and the parameters wander near the 
global minimum of $\chi^2$ in almost a flat valley of some complicated 
topology. For example, the best-fitting flat model 
($\Omega_{\Lambda0}=1-\Omega_{\rm m0}$) is obtained as 

\begin{center}
\begin{tabular}{llll}
$\Omega_{\rm m0}=0.62\pm0.03$, &   
$dh_0=6.46\pm0.31$, &
$\chi^2/$DoF$=13.09/14$,&  
$Q=51.9\%$, 
\end{tabular}
\end{center}
which represents a slightly better fit (as the number of DoF is increased). It may be mentioned that the value 
$\Omega_{\rm m0}=0.62$ estimated from these data is in good agreement
with the results estimated (in the following section) from the recent 
observations of SNe 
of type Ia. It should be noted that the best-fitting solution 
$\Omega_{\rm m0}=0.62$ gives a mildly accelerating expansion of the universe at the present epoch: the deceleration parameter $q_0\equiv\Omega_{\rm m0}/2-\Omega_{\Lambda0}=-0.07$.
In order to compare, we find that the best-fitting concordance model
(flat standard $\Lambda$CDM model with a constant $\Lambda$)
 to this dataset is obtained as

\begin{center}
\begin{tabular}{llll}
$\Omega_{\rm m0}=0.21\pm0.08$,&  
$dh_0=7.25\pm0.55$,&
$\chi^2/$DoF$=16.03/14$,& 
$Q=31.1\%$, 
\end{tabular}
\end{center}
which also represents a good fit. 
One may note that the  value $\Omega_{\rm m0}=0.62$ obtained for the model 
$\rho_\Lambda \propto S^{-2}$ is higher than $\Omega_{\rm m0}=0.21$ obtained for the standard 
$\Lambda$CDM model. However, one should note that the other precise observations, which one 
would expect to be consistent with any model, are the measurements of the temperature anisotropy 
of CMB made by the WMAP experiments \cite{wmap}, whose only apparent prediction is  
$\Omega_{\rm m0}+\Omega_{\Lambda0}=1$ \cite{blanchard}. For this reason, and 
also motivated by theoretical considerations required by inflation and flatness problem, we assume spatial flatness henceforth.  
For curiosity, we test the Einstein-de Sitter model ($\Omega_{\rm m0}=1$, 
$\Lambda=0$) against the radio sources data: the best fitting solution is 
obtained as 

\begin{center}
\begin{tabular}{llll}
$dh_0=4.91\pm0.11$,&  
$\chi^2/$DoF$=28.82/15$,&
$Q=1.69\%$, 
\end{tabular}
\end{center}
which can be rejected by the data only at 98.3\% confidence level. 
In Figure 1, we have shown some models obtained from our fitting procedure
and compared them with the data.

\section{Fitting the model to the recent SNe Ia data}

\begin{figure}[tbh!]
\centerline{{\epsfxsize=14cm {\epsfbox[50 250 550 550]{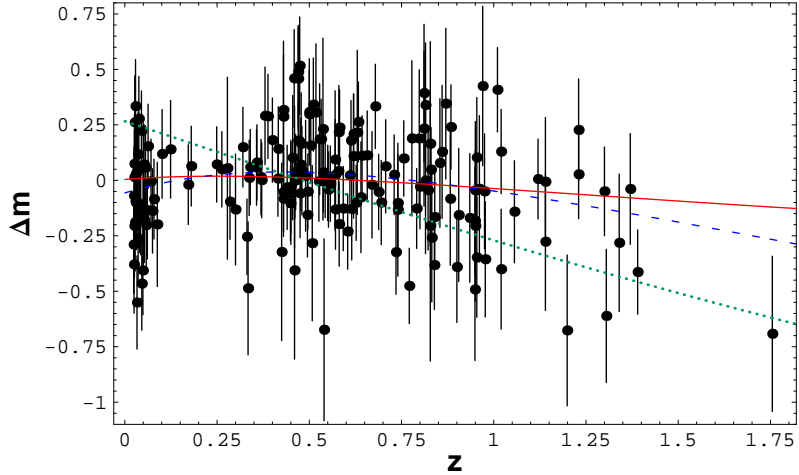}}}}
{\caption{\small Modified Hubble diagram of new `gold sample' of SNe Ia 
\cite{riess} minus a 
fiducial model ($\Omega_{\rm m0}=\Omega_{\Lambda0}=0$): 
The relative magnitude($\Delta m \equiv m-
m_{\rm fiducial}$) is plotted for some best-fitting models.
The solid curve corresponds to the dark energy model 
$\rho_{\Lambda}\propto S^{-2}$ with $\Omega_{\rm m0}=1-\Omega_{\Lambda0}=0.58$, the dashed  
curve corresponds to the standard concordance model with $\Omega_{\rm m0}=1-\Omega_{\Lambda0}=0.34$
and the dotted curve corresponds to the Einstein-de Sitter model. 
}}
 \end{figure}

\noindent
It was already shown that the model explained the data of SNe of type Ia from
Perlmutter et al. successfully \cite{perl}. Since then many SNe of type Ia at
higher redshifts have been discovered. Extending our earlier work further, we
now examine how the model fits the updated \emph{gold sample} of 
Riess et al.
\cite{riess}. In addition to having previously 
discovered SNe Ia, this sample (of 182 SNe Ia in total) contains 23 SNe Ia at $z\geq1$ recently
discovered by the Hubble Space Telescope and is claimed to have a high-confidence-quality of
spectroscopic and photometric record for the individual SNe. 

We find that the present model has an excellent fit to this data, comparable
with the concordance model. The best-fitting concordance model is obtained as

\begin{center}
\begin{tabular}{llll}
$\Omega_{\rm m0}=0.34\pm0.04$,&  
${\cal M}=43.40\pm0.03$,&
$\chi^2/$DoF$=158.75/180$,& 
$Q=87.1\%$, 
\end{tabular}
\end{center}
an excellent fit indeed! The model $\Lambda \propto S^{-2}$ 
provides a similar fit:

\begin{center}
\begin{tabular}{llll}
$\Omega_{\rm m0}=0.58\pm0.02$,&  
${\cal M}=43.46\pm0.03$,&
$\chi^2/$DoF$=167.90/180$,& 
$Q=73.2\%$. 
\end{tabular}
\end{center}
The Einstein-de Sitter model does not fit the data well:
\begin{center}
\begin{tabular}{lll}
${\cal M}=43.72\pm0.02$,&  
$\chi^2/$DoF$=283.40/181$,&
$Q=1.7\times10^{-4}\%$. 
\end{tabular}
\end{center}
These models have been shown in Figure 2. In order to have a visual comparison of the fits of different models to the
actual data points, we
magnify their differences by plotting the relative magnitude with respect to
a fiducial model $\Omega_{\rm m0}=\Omega_{\Lambda0}=0$ (which also has a good 
fit: $\chi^2$/DoF $=174.29/181$, $Q=62.6\%$).

\section{Conclusions}

\noindent
In order to test the consistency of the cosmological models with
observations as well as to estimate the different cosmological
parameters, data on SNe of type Ia and radio sources have been used by several
authors.
We use the recent gold sample of SNe Ia from Riess et al. and 
the sample of 256 ultracompact radio
sources (of angular sizes of the order of a few milliarcseconds) compiled by Jackson and Dodgson
to test a model of dark energy which results consistently from the contracted
Ricci-collineation along the fluid flow vector. We find that the model has excellent fits to both
the data sets and the estimated parameters are also in good agreement.

\section{Acknowledgement} The author thanks Abdus Salam ICTP for hospitality.

\bigskip
\noindent
{\bf APPENDIX}

\setcounter{equation}{0}
\renewcommand{\theequation}{A.\arabic{equation}}

Though there is not a clearly defined value of $\chi^2$/DoF for an acceptable
fit, however
it is obvious from equation (\ref{eq:chi}) that if the model represents
the data correctly, the difference between the predicted angular 
size/magnitude and 
the observed one at each data point should be roughly the same size as the 
measurement uncertainties and each data point will contribute to $\chi^2$ 
roughly one, giving the sum roughly equal to
the number of data points $N$ (more correctly $N-$number of fitted parameters
$\equiv$ number of degrees of freedom `DoF'). This is regarded as
a \emph{`rule of thumb'} for a \emph{moderately} good fit. 
If $\chi^2$ is large, the fit is bad. However we must quantify our 
judgment and decision about the \emph{goodness-of-fit}, in the absence of 
which, the estimated parameters of the model (and their estimated 
uncertainties) have no meaning
at all. An independent assessment of the goodness-of-fit of the data to the 
model is given in terms of the $\chi^2$-\emph{probability}:
if the fitted model
provides a typical value of $\chi^2$ as $x$ at $n$ DoF, this probability is
given by
\begin{equation}
Q(x, n)=\frac{1}{\Gamma (n/2)}\int_{x/2}^\infty e^{-u}u^{n/2-1} {\rm d}u.
\end{equation}
Roughly speaking, it measures \emph{the probability that the model does
describe the data genuinely and any discrepancies are mere fluctuations which 
could have arisen by chance}. To be more precise, $Q(x, n)$ gives the probability that a model
which does fit the data at $n$ DoF, would give a value of $\chi^2$ as large
or larger than $x$. If $Q$ is very small, the apparent discrepancies are
unlikely to be chance fluctuations and the model is ruled out. 
For example, if we get a $\chi^2=20$ at 5 DoF for some model, then the 
hypothesis that {\it the model describes the data genuinely} is  unlikely, 
as the probability $Q(20, 5)=0.0012$ is very small.
It may however
be noted that the $\chi^2$-probability strictly holds only when the models are
linear in their parameters and the measurement errors are normally distributed.
It is though common, and usually not too wrong, to assume that the 
$\chi^2$- distribution holds even for 
models which are not strictly linear in their parameters, and for this reason,
the models with a probability as low as $Q>0.001$ are usually deemed 
acceptable \cite{press}.

\end{document}